\documentclass[sigconf]{acmart}

\usepackage{booktabs,multirow,mathtools}
\usepackage{algorithm,algcompatible} 

\usepackage{enumitem}

\begin{document}
\title{PS-DBSCAN: A Communication Efficient Parallel DBSCAN Algorithm Based on Platform Of AI (PAI)}


\author{Xu Hu}
\affiliation{%
	\institution{Alibaba Group}
	\streetaddress{}
	\city{} 
	\state{} 
	\postcode{}
}
\email{huxu.hx@alibaba-inc.com}

\author{Jun Huang}
\affiliation{%
  \institution{Alibaba Group}
  \streetaddress{}
  \city{} 
  \state{} 
  \postcode{}
}
\email{huangjun.hj@alibaba-inc.com}

\author{Minghui Qiu}
\affiliation{%
  \institution{Alibaba Group}
  \streetaddress{}
  \city{}
  \state{}
  \postcode{}
}
\email{minghui.qmh@alibaba-inc.com}

\author{Cen Chen}
\affiliation{%
  \institution{Ant Financial Services Group}
  \streetaddress{}
  \city{}
  \state{}
  \postcode{}
}
\email{chencen.cc@antfin.com}

\author{Wei Chu}
\affiliation{%
  \institution{Alibaba Group}
  \streetaddress{}
  \city{}
  \state{}
  \postcode{}
}
\email{weichu.cw@alibaba-inc.com}

\begin{abstract}
We present PS-DBSCAN, a communication efficient parallel DBSCAN algorithm that combines the disjoint-set data structure and Parameter Server framework in Platform of AI (PAI). Since data points within the same cluster may be distributed over different workers which result in several disjoint-sets, merging them incurs large communication costs. In our algorithm, we employ a fast global union approach to union the disjoint-sets to alleviate the communication burden. Experiments over the datasets of different scales demonstrate that PS-DBSCAN outperforms the PDSDBSCAN with 2-10 times speedup on communication efficiency.

We have released our PS-DBSCAN in an algorithm platform called Platform of AI (PAI)~\footnote{PAI: Platform of Artificial Intelligence \url{https://pai.base.shuju.aliyun.com/}} in Alibaba Cloud. We have also demonstrated how to use the method in PAI.
\end{abstract}

%
%

%

\keywords{Density-Based clustering, Parallel DBSCAN, Parameter Server}

\maketitle

\section{Introduction}
Clustering is an unsupervised data mining technology that divides a set of objects into subgroups by maximizing inter-group distances and minimizing intra-group distances. 
Usually, the clustering algorithm can be divided into four classes: partition-based, hierarchy-based, grid-based and density-based.  
Among all the clustering algorithms, DBSCAN~\cite{dbscan96KDD}, a density-based algorithm, is one of the most popular.
The key idea of DBSCAN is that, for one point $p$ of the data set in d-dimensional space $\mathbb{R}^d$, if its neighborhood within the d-dimensional ball with radius $\epsilon$, i.e., $\epsilon$-neighborhood, contains at least $minPoints$ points, all the points inside this ball including $p$ formed a cluster. And $p$ is defined as a core point. Whenever a new core point is added to the cluster of $p$, all the points within the new core point's $\epsilon$-neighborhood are added to the cluster. This process goes on recursively in this way until all the clusters extended to their maximum size. 

DBSCAN is a computationally expensive algorithm, with $O(n^2)$ time complexity as shown in ~\cite{dbscanrevisited}, this makes it inefficient for clustering large scale data sets. Thus we focus on parallelizing DBSCAN in this study.
Recently, a benchmark work compared the performance of the current parallelized implementations in terms of runtime and found that MPI based implementations are more efficient than the others~\cite{neukirchen2016survey}. In a typical MPI implementation of DBSCAN such as PDSDBSCAN-D, data points are partitioned uniformly into different processors, and core points and their $\epsilon$-neighborhood points may be distributed among processors, thus communications are required to merge these points to one cluster. When the number of processors increases, the message number and communication frequency will increase and the communication time will be dominant~\cite{Patwary:2012}. 

\begin{figure}[t!]
	\centering
	\includegraphics[width=1.0\columnwidth]{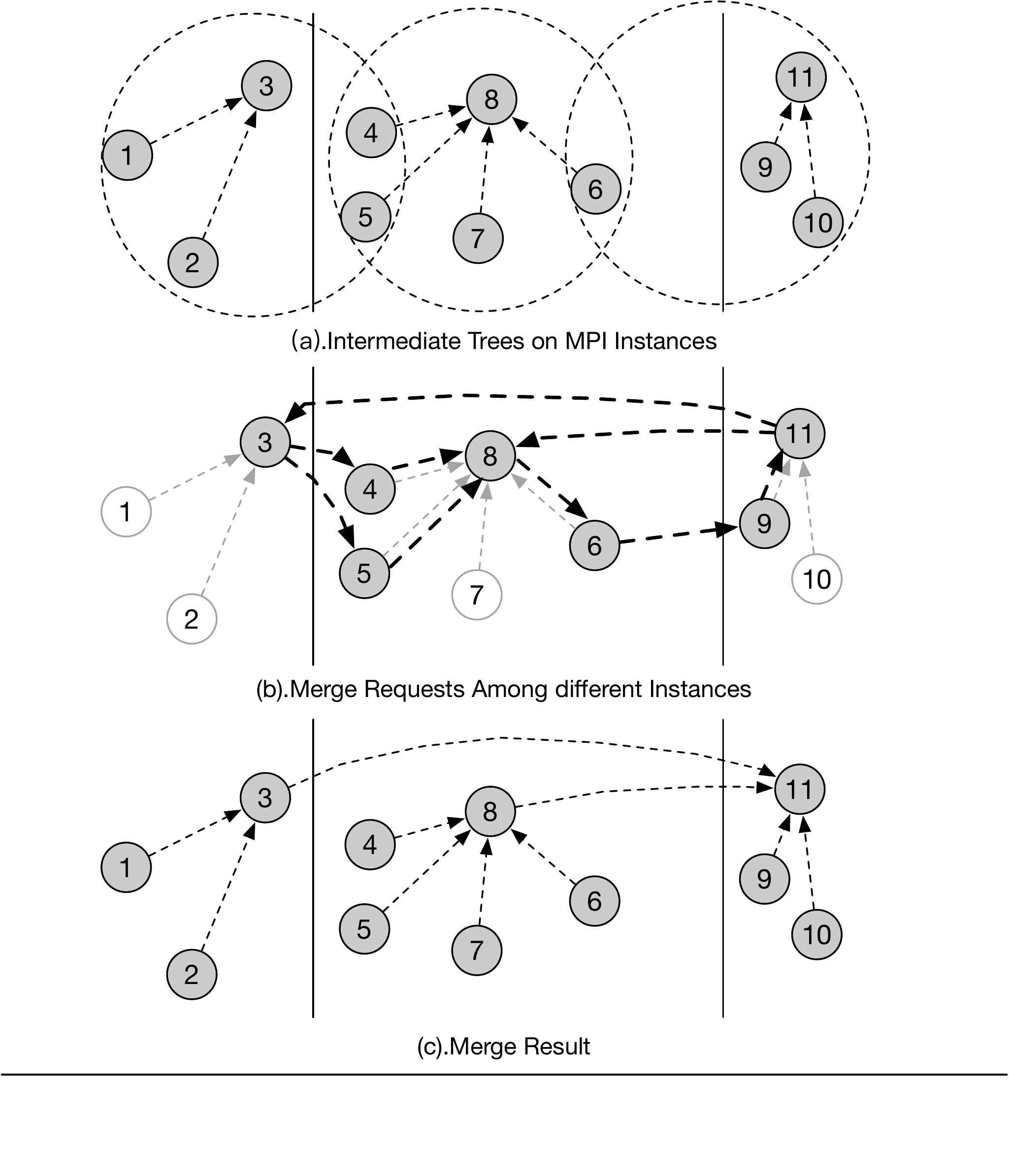}
	\vspace{-8mm}
	\caption{The communication mode of MPI based DBSCAN.}
	\label{fig:mpi}
\end{figure}
Fig.~\ref{fig:mpi} illustrates the communication mode of typical MPI based DBSCAN. As in Fig.~\ref{fig:mpi}(a), all the core points that form a single cluster are distributed over three workers. Clearly, through common neighbor points $4$ and $5$, node $3$ needs to route from node $8$ and $6$ to reach its parent node $11$. Since MPI uses a peer-to-peer communication pattern, this process will generate a lot of merging requests. In general, the MPI based setting will have communication overhead when points from the same cluster are scattered over more partitions. And this scenario will be worse with the increase of worker number. More details of the communication process are in\cite{Patwary:2012}.

To overcome the communication bottleneck, we employ a parameter server framework~\cite{li2014scaling} to implement parallel DBSCAN algorithm using the disjoint-set data structure mentioned in the paper ~\cite{Patwary:2012}. The details about our Parameter Server framework can be found in~\cite{cc-www17,alips-kdd17}. 
In our proposed algorithm, a global vector that records the class label of all data points is stored in the server processors. In worker processors, we employ a fast global union approach to union the disjoint-sets locally and push the resulted label vector to servers to update the global vector. This method alleviates the communication burden. Experiments over the datasets of different scales demonstrate that PS-DBSCAN outperforms PDSDBSCAN-D with 2-10 times speedup on communication efficiency.

The remainder of this paper is organized as follows. Section 2 describes the details of our parallel implementation of DBSCAN based on Parameter Server framework, referred to as PS-DBSCAN. In section 3 we compare the speedup of communication between our algorithm and the MPI based method PDSDBSCAN-D. Section 4 demonstates the usage of our PS-DBSCAN in our PAI. In section 5, we survey the related work. Section 6 gives a brief conclusion and an overview of future work.

\begin{figure}[th!]
	\centering
	\includegraphics[width=1.0\columnwidth]{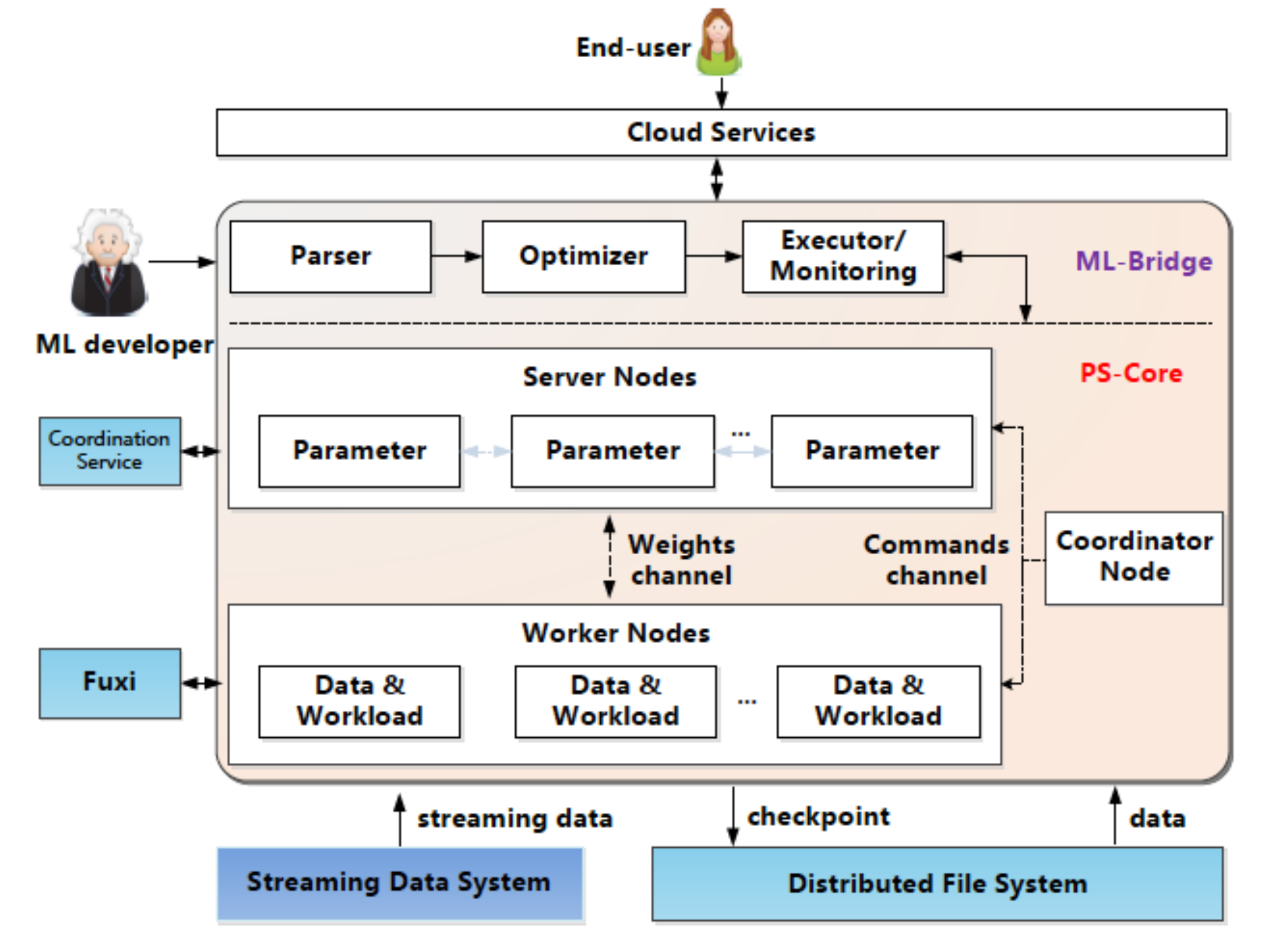}
	\caption{Kunpeng system~\cite{alips-kdd17}.}
	\label{fig:Kunpeng}
\end{figure}

\section{Methodology}
Our PS-DBSCAN is built based on Alibaba parameter server system called KunPeng~\cite{alips-kdd17}. The KunPeng architecture is shown in~Fig.~\ref{fig:Kunpeng}. We use SDK of KunPeng to implement the distributed algorithm.

\begin{figure*}[th!]
	\centering
	\includegraphics[width=1.8\columnwidth]{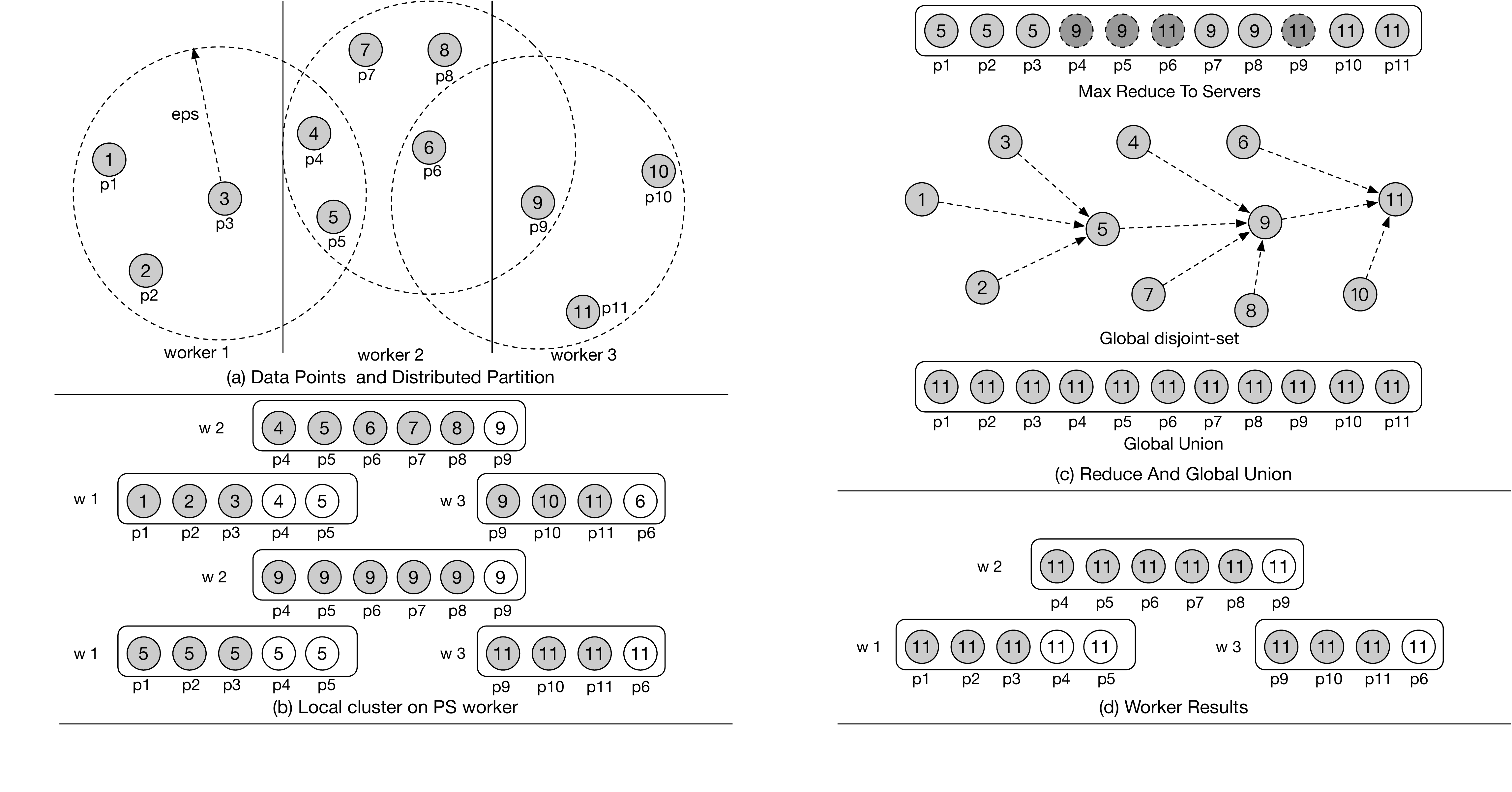}
	\vspace{-6mm}
	\caption{Sample workflow of PS-DBSCAN.}
	\label{fig:example}
\end{figure*}
\begin{figure}[th!]
	\centering
	\includegraphics[width=1.0\columnwidth]{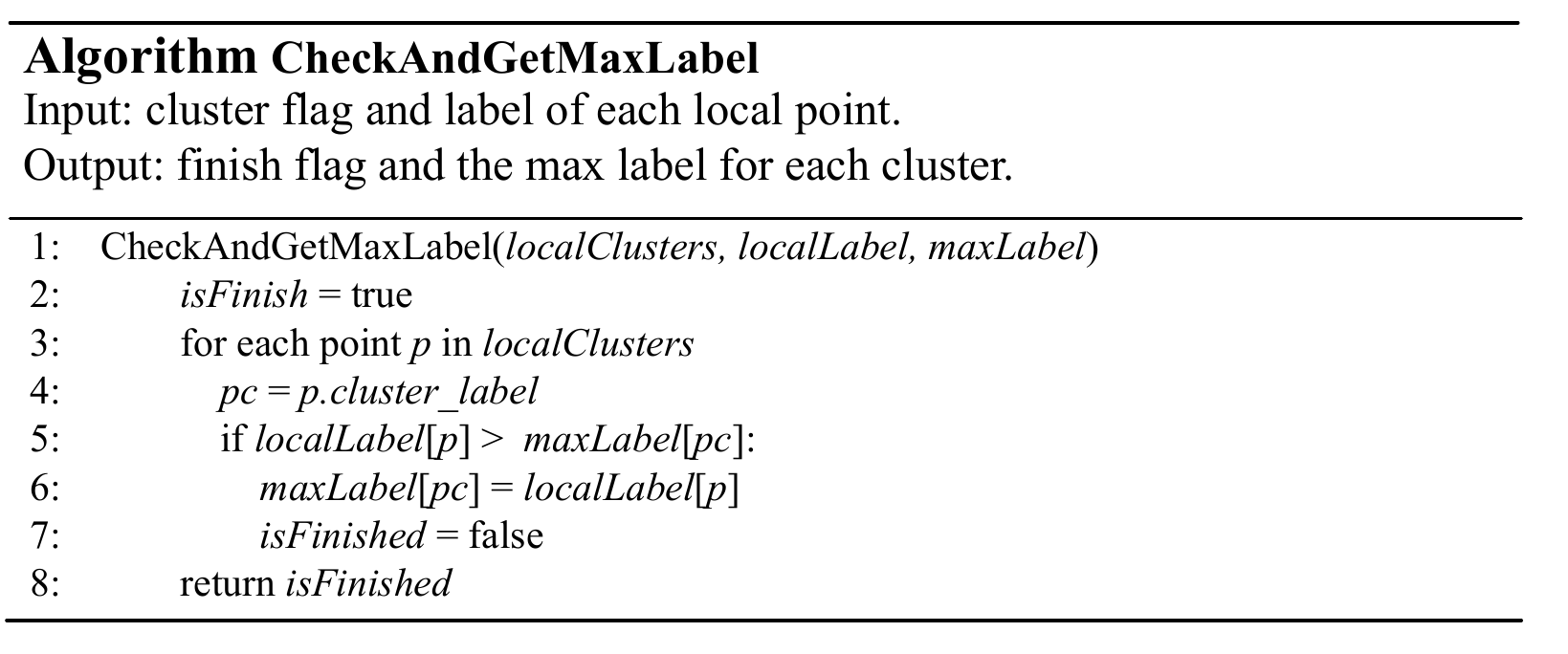}
 	\caption{Peseudocode of CheckAndGetMaxLabel.}
 	\label{fig:union}
 \end{figure}

To illustrate our algorithm, we use Fig.~\ref{fig:example} as a running example. 
Our algorithm starts by randomly dividing the input data points $Pts$ into $p$ partitions and distributing them to $p$ workers, e.g., in Fig.~\ref{fig:example}(a), nodes $1\sim 3$ are in worker $w_1$ and nodes $4\sim 8$ are in worker $w_2$. In our setting, we have servers to maintain $globalLabel$, and local workers to maintain their own $localLabel$. Initially, all the workers perform clustering operations in parallel, where each worker uses $QueryRadius$ to find each local data point's $\epsilon$-nearest neighbors and $MarkCorePoints$ accordingly. All the $localCoreRecord$ will be synchronized with the servers to get $globalCoreRecords$. A $LocalMerge$ operation is performed by each worker to create $localCluster$ based on the $\epsilon$-nearest neighborhood information and $globalCoreRecord$. With the $localCluster$, all the workers start to label its local data points and communicate with servers to remove labeling conflicts. The steps $PropagateMaxLabel$, $MaxReduceToServer$, $PullFromServer$, $GlobalUnion$, and $GetMaxLabel$ are performed iteratively until no labeling conflicts found.
The key steps are discussed as follows.
\begin{itemize}[leftmargin=5mm]
    \item MarkCorePoint: A point $p$ is marked as a core point if its $\epsilon$- neighborhood size is at least $minPoints$. 
    \item PropagateMaxLabel: This is a local clustering processing where all the nodes in the same cluster are labeled as the maximum local node id. As in Fig.~\ref{fig:example}(b), node $4\sim 9$ are labeled with id $9$.
    \item MaxReduceToServer: A Synchronous Max Reduce operator is used to merge local clustering results with server results, where each node will be labeled as the maximum node id from all local workers. As in Fig.~\ref{fig:example}(c), node $6$ takes $11$ from $w_3$, i.e. $\max(9|w_2,11|w_3)$.
    \item PullFromServer: This is a typical PS operator to pull results from the server. Interested readers can refer to~\cite{li2014scaling} for details.
    \item GlobalUnion: This step starts from the maximum node id $N-1$ to 0, for each node, if its root node id does not equal to the corresponding global label, we modify it to the global label. This is an effective way to compress the path of disjoint-set and redirect each local node to its root parent. For example, in Fig.~\ref{fig:example}(c), node $3$ will directly link to its root node $11$. Unlike Fig.~\ref{fig:mpi}, where node $3$ needs to route from nodes $8$ and $6$ to link to $11$. This is the key step to reduce communication burden. 
    \item GetMaxLabel: This step is performed on the local cluster to label each data point with the maximum node id within a cluster. The detailed algorithm is described as in Fig.~\ref{fig:union}. After this step, all the local nodes are labeled as the maximum node within the cluster. As shown in Fig.~\ref{fig:example}(d), with this step, all the local nodes in $w1$ are labeled as node 11.
\end{itemize}

We present our PS-DBSCAN method in Algorithm~\ref{alg:dbscan}. 
\begin{algorithm}                   
\caption{PS-DBSCAN}          
\label{alg:dbscan}                           
\begin{algorithmic}[1]                    
    \STATE \textbf{Input:} A set of points $Pts$, distance threshold $\epsilon$ and density threshold $minPoints$.
    \STATE \textbf{Output:} clusters of data points
    \STATE {Randomly divide $Pts$ into partitions and distribute to workers.}
    \STATE {InitOnServer($globalCoreRecord$,$globalLabel$)}
    \STATE InitOnWorker($localCoreRecord$,$localLabel$,$localCluster$)
    \FOR{each worker $i$, parallel do}
        \FOR{each point $p$ in $Pts_i$}
            \STATE{$p_n$ = QueryRadius($p$,$\epsilon$)}
            \STATE{$localCoreRecord$ = MarkCorePoint($p_n$,$p$,$minPoints$)}
        \ENDFOR
        \STATE{ReduceToServer($localCoreRecord$)}
        \STATE{PullFromServer($globalCoreRecord$)}
        \STATE{$localCluster$ = LocalMerge($p_n$,$globalCoreRecord$)}
        \STATE{$isFinish,maxLabel$ = GetMaxLabel($localClusters,localLabel$)}
        \WHILE{not $isFinish$}
            \STATE {PropagateMaxLabel($localClusters,localLabel,maxLabel$)}
            \STATE{MaxReduceToServer($localLabel$)}
            \STATE{PullFromServer($globalLabel$)}
            \STATE{GlobalUnion($localLabel$)}
            \STATE{$isFinish,maxLabel$ = GetMaxLabel($localClusters,localLabel)$}
        \ENDWHILE
    \ENDFOR
    \STATE \textbf{Return:} $globalLabel$
\end{algorithmic}
\end{algorithm}

In a nutshell, comparing with the MPI-based PDSDBSCAN-D, our method has two advantages. First, each worker maintains a local cluster and we only generate merging requests when it has modified labels. This can help to reduce communication overhead. Second, with $GlobalUnion$, each data point is able to find its root parent directly without generating many merge requests. This makes our algorithm 2-10 times faster than the PDSDBSCAN-D.


\section{Experiments}
We quantitatively evaluated our PS-DBSCAN here. We first designed experiments to examine the communication efficiency and speedup gain of our method comparing to the MPI-based PDSDBSCAN. Our method has better scalability than PDSDBSCAN where it shows good performance with up to 1600 CPU cores. 

\noindent \textbf{Setup.} We evaluated our methods on a cluster where each computer node has 24 cores, 4 Intel Xeon E5-2430 hex-core processors, and 96GB memory. We implemented the PDSDBSCAN-D with open source code~\footnote{\url{http://cucis.ece.northwestern.edu/projects/Clustering/}} on the cluster. As only single-threaded implementation of PDSDBSCAN-D is available, we limited to use one core in each computer node in our experiments. 
Note that, the cluster is used as a production cluster shared by many applications, to avoid the impact of other tasks, we repeated the experiments 6 times and take the mean results by ignoring the best and worst results.

\noindent \textbf{Datasets.} To investigate the performance of our PS-DBSCAN, we first generated two \textit{synthetic datasets}: $D10m$ and $D100m$. $D10m$ has 10 million data points and each data point has an average of 25 directly density-reachable core points (or $\epsilon$-neighborhood), while $D100m$ has 100 million points and each has 15 $\epsilon$-neighborhood. We pre-computed pair-wise distance information for both of them. 

Furthermore, we used two large \textit{real-world datasets} from~\cite{hpdbscan:15}, one is Geo-tagged tweets, and the other BremenSmall that contains 3D-point cloud of an old town. The Tweets was obtained using the free twitter streaming API and contains location of all geo-tagged tweets, it consists of 16,602,137 2D-points. And BremenSmall is a set of 3D-point cloud of the old town of Bremen, which contains 2,543,712 points.


\subsection{Examination of Communication Efficiency}
 Table~\ref{tab:all} shows the communication time of MPI-based PDSDBSCAN-D and our PS-DBSCAN on synthetic and real-word datasets using 100,200,400,800 and 1600 cores. Some important observations are discussed in order. 
 
 First, on all the datasets, the PDSDBSCAN-D tends to be slower than our PS-DBSCAN with the increase of CPU nodes. The reason is that PDSDBSCAN's peer-to-peer communication pattern has communication overhead with a large number of CPU nodes.
 
 Second, our PS-DBSCAN has a very limited number of communication iterations regardless of the growing number of CPU nodes. This is because our global union methods help to reduce the number of merging requests.
 
 Third, MPI-based PDSDBSCAN-D is not stable with a large number of CPU nodes. For example, with 1600 CPU nodes, PDSDBSCAN fails to generate results, while our PS-DBSCAN still works. Furthermore, the PDSDBSCAN is severely affected by a large amount of the neighbors. For Tweets datasets with 169 $\epsilon$-nearest neighbors when $\epsilon=0.01$ and 3600 neighbors when $\epsilon=0.01$, PDSDBSCAN fails. Both of these problems make PDSDBSCAN not ideal for a very large data set.
 
 Last but not least, on the largest dataset $D100m$, the communication time of PS-DBSCAN decreases first and then increases as the nodes increases.
 Close examination shows, when the amount of the data points is too large, the total merge time will benefit from the increase in the number of nodes to some extent.

\begin{table}[th!]
	\caption{Communication time on all datasets.}
	\vspace{-3mm}
	\label{tab:all}
	\begin{tabular}{l r r r r r}
		\toprule
		Cores & 100 & 200 & 400 & 800 & 1600\\
		\midrule
		\multicolumn{6}{c}{$D10m$ (125million edges)}\\
		\hline
		PDSDBSCAN-D & 37.52 & 51.34 & 102.78 & 120.23 & NA\\
		PS-DBSCAN & 9.23 & 10.18 & 11.12 & 11.4 & 24.78\\
		Speedup & 4.07x & 5.04x & 9.24x & 10.55x & \\
		\hline
		\multicolumn{6}{c}{$D100m$ (750million edges)}\\
		\hline
        PDSDBSCAN-D & 243.44 & 202.23 & 204.64 & 263.34 & NA\\
		PS-DBSCAN & 71.81 & 56.18 & 39.24 & 46.54 & 52.83\\
		Speedup & 3.39x & 3.60x & 5.22x & 5.66x & \\
		\bottomrule
		\multicolumn{6}{c}{BremenSmall ($\epsilon$=10,points=2,543,712)}\\
		\hline
		PDSDBSCAN-D & 17.15 & 48.17 & 61.08 & 70.11 & NA\\
		PS-DBSCAN & 8.5 & 14.26 & 14.86 & 15.02 & 20.64\\
		Speedup & 2.01x & 3.38x & 4.11x & 4.67x & \\
		\hline
		\multicolumn{6}{c}{Tweets ($\epsilon$=0.000001,points=16,602,137)}\\
		\hline
        PDSDBSCAN-D & NA & NA & NA & NA & NA\\
		PS-DBSCAN & 13.04 & 13.31 & 17.74 & 18.95 & 21.16\\
		\hline
		\multicolumn{6}{c}{Tweets ($\epsilon$=0.01,points=16,602,137)}\\
		\hline
        PDSDBSCAN-D & NA & NA & NA & NA & NA\\
		PS-DBSCAN & 23.75 & 26.52 & 26.7 & 29.79 & 36.16\\
		\bottomrule
	\end{tabular}
\end{table}

 
\subsection{Examination of Speedup Gains}
 We further examined the speedup gains of our PS-DBSCAN over PDSDBSCAN-D.
 
\begin{figure}[th!]
	\centering
	\includegraphics[width=1.0\columnwidth]{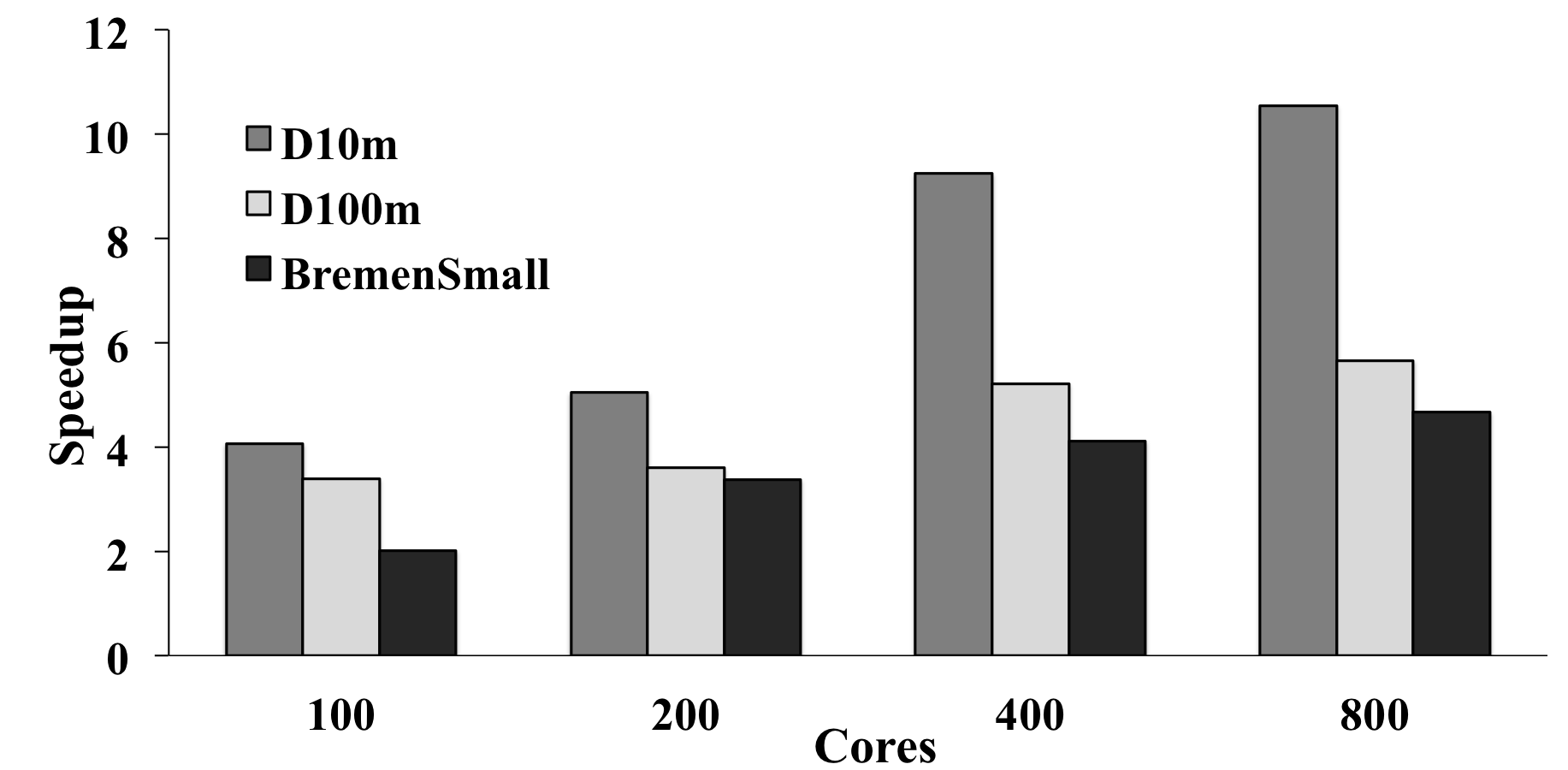}
	\caption{PS-DBSCAN speedup on different datasets.}
	\label{fig:speedup}
\end{figure}

 As in Fig.~\ref{fig:speedup}, with more CPU cores, our method has a larger speedup gain. In general, PS-DBSCAN outperforms the PDSDBSCAN with 2-10 times speedup on communication efficiency.
\begin{figure}[th!]
	\centering
	\includegraphics[width=1.0\columnwidth]{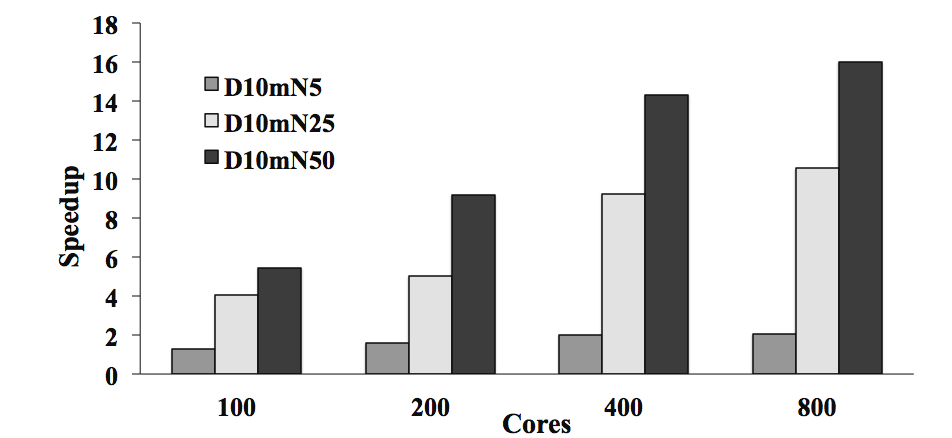}
	\caption{PS-DBSCAN speedup on $D10m$.} 
	\label{fig:speedup2}
\end{figure}

Particularly, we found PS-DBSCAN has 10 times speedup with 800 CPUs nodes on $D10m$, which is significantly larger than on other datasets. Close examination shows that MPI-based DBSCAN suffers from a large $\epsilon$-nearest neighborhood size. To illustrate this, we used three datasets $D10mN5$, $D10mN25$ and $D10mN50$, corresponding to a neighborhood size of $5$, $25$ and $50$ respectively, to evaluate their performance in Fig.~\ref{fig:speedup2}. Clearly, PDSDBSCAN has a degenerated performance with a larger neighborhood size. The reason is that with a larger neighborhood size, each core point has more neighbors being distributed to different workers which result in generating more merging requests in MPI setting. While in PS-DBSCAN, with maintaining a global label and using $GlobalUnion$, there are far fewer merging requests.

We have released our PS-DBSCAN in an algorithm platformcalled Platform of AI (PAI) in Alibaba Cloud. Below we demonstrate the usage of PS-DBSCAN in our cloud-based platform - PAI.

\begin{figure*}[th!]
	\centering
	\includegraphics[width=1.9\columnwidth]{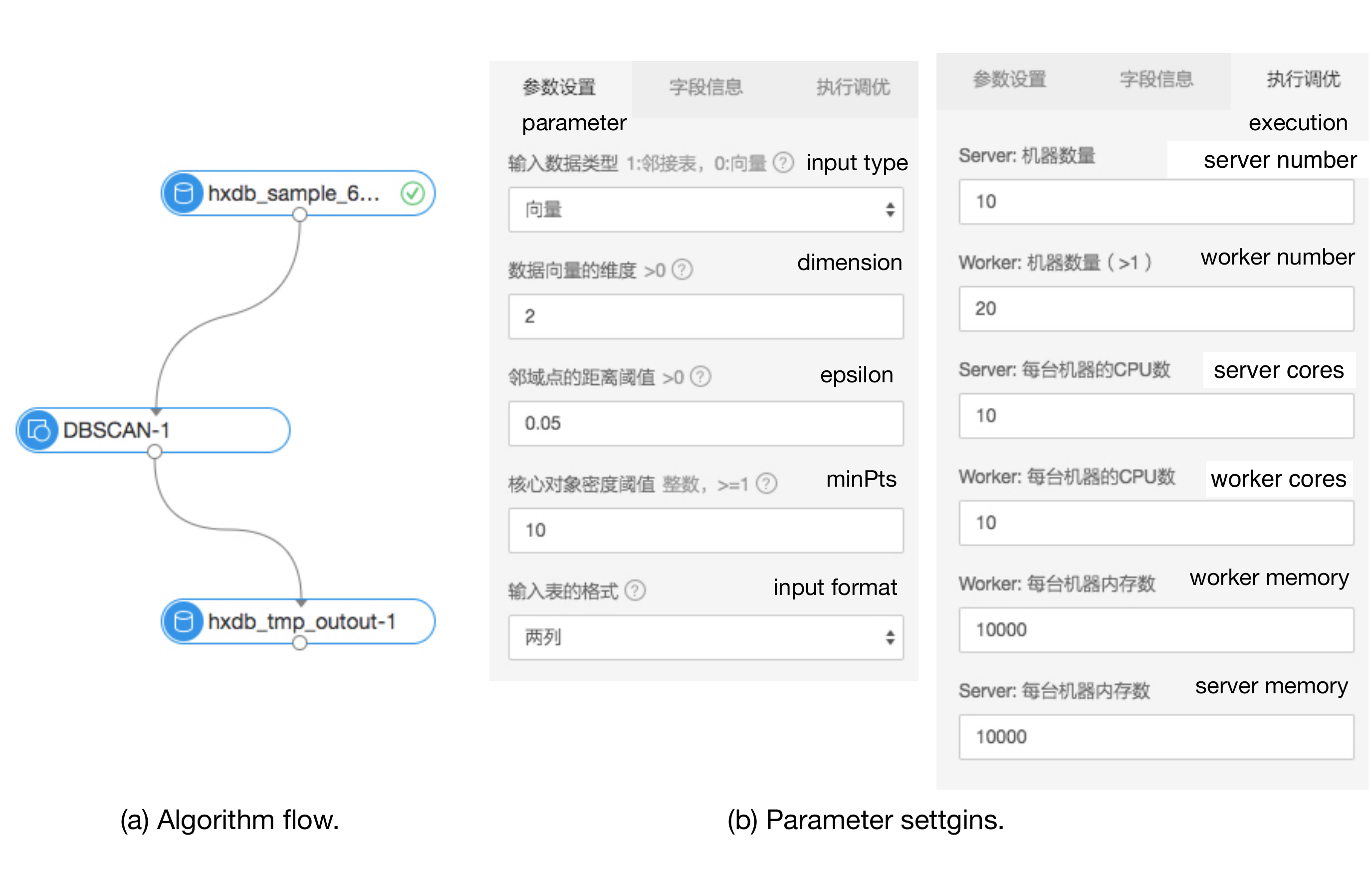}
	\caption{PS-DBSCAN component in PAI.} 
	\label{fig:usage}
\end{figure*}
\section{Demonstration}
In this section, we demonstrate the usage of PS-DBSCAN in PAI. PAI provides an interface to interact with PS-DBSCAN component. The whole workflow is shown in Fig~\ref{fig:usage}(a), where an input table named as ``hxdb\_sample\_6" is linked to the PS-DBSCAN component ``DBSCAN-1". The output of the component is linked to an output table ``hxdb\_tmp\_output-1". With this workflow, the method automatically pulls the data from the input table and run the PS-DBSCAN algorithm, and the final results are stored in the output table.

We also provide an interface for users to tune the parameters, as in Fig~\ref{fig:usage}(b). Specifically, we can tune the following parameters based on the interface.
\begin{itemize}
    \item Input type: vector or linkage
    \item Dimension: input data dimension
    \item Epsilon: the distance threshold of DBSCAN
    \item minPts: the density threshold of DBSCAN
    \item input format: the number of input columns
    \item server number: the number of server nodes
    \item worker number: the number of worker nodes
    \item server cores: CPU cores for each server
    \item worker cores: CPU cores for each worker
    \item server memory: server memory
    \item worker memory: worker memory
\end{itemize}

\begin{figure}[th!]
	\centering
	\includegraphics[width=1.0\columnwidth]{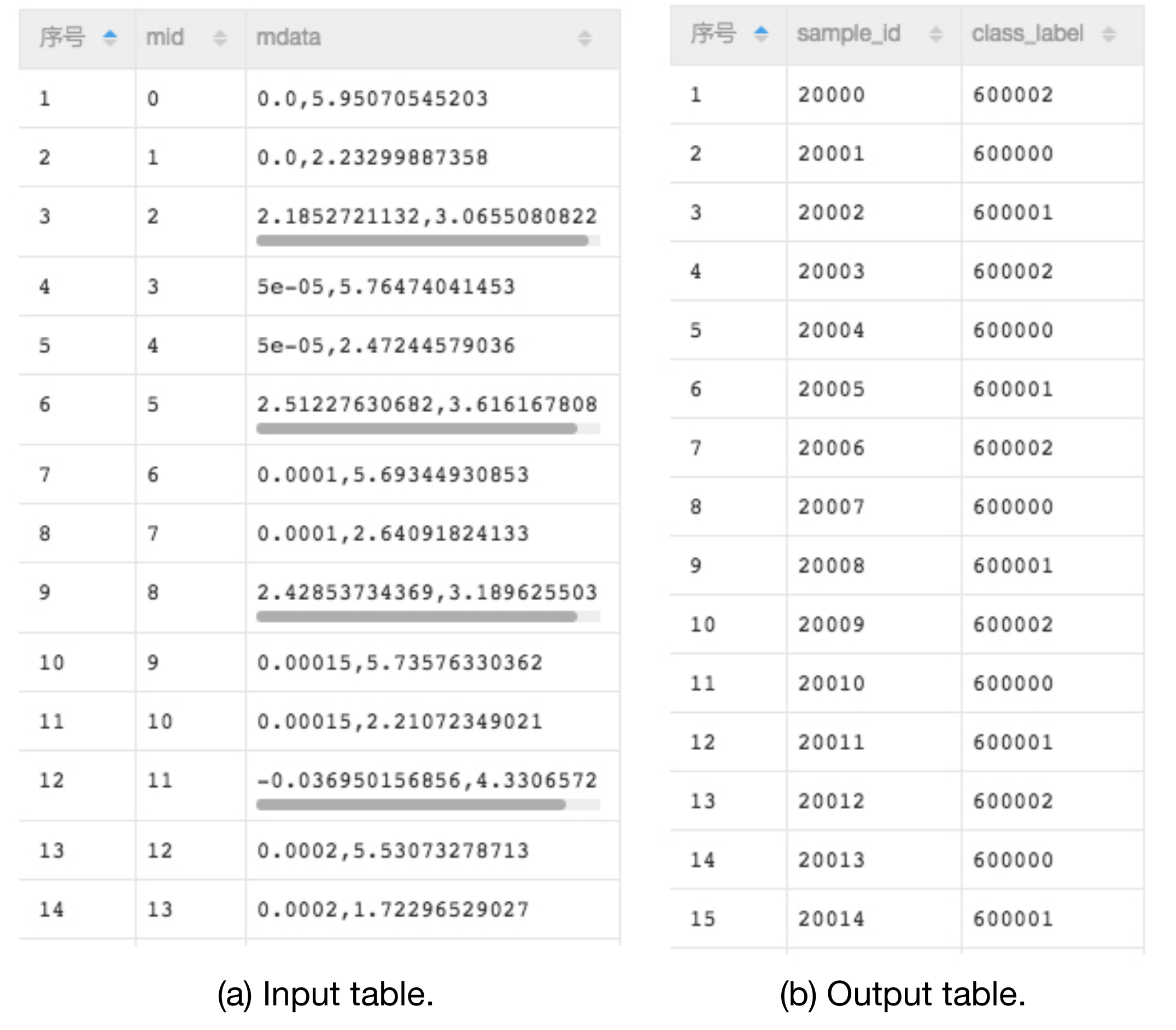}
	\caption{Input and output tables of the PS-DBSCAN algorithm in PAI.} 
	\label{fig:IO}
\end{figure}
We present the input and output tables of our PS-DBSCAN algorithm in Fig~\ref{fig:IO}. The table is stored in MaxCompute platform. Interested readers can find the details here: \url{https://www.aliyun.com/product/odps/}.

We support two types of data as input:
\begin{itemize}
    \item Vector: each node has an index and is represented by a vector, as shown in Fig~\ref{fig:IO}(a).
    \item Linkage: each record in the table is a link between two nodes.
\end{itemize}

After running this algorithm, we can get the clustering result of our input data, as shown in Fig~\ref{fig:IO}(b).

To test the PS-DBSCAN method, users can register PAI online via this link~\url{https://pai.base.shuju.aliyun.com/} and search for PS-DBSCAN in the search bar.

\section{Related Work}
There are generally two lines of work for paralleling DBSCAN, one is on MapReduce-based big data platforms such as Apache Spark and the other is on distributed memory using Message Passing Interface-based (MPI).

The studies in~\cite{fu:2011,He:2011} are the first to implement a  parallel DBSCAN based on the Map-Reduce paradigm. A similar idea is used in RDD-DBSCAN~\cite{Irving:15}.
In~\cite{Irving:15,Litouka}, the data space is split into roughly equal sized boxes until the data size of a box is less or equal to a threshold, or a maximum number of levels is reached, or the shortest side of a box becomes smaller than 2 eps. Each resulting box is a record of an RDD which can be processed in parallel. Another work~\footnote{Mansour Raad. 2016.  \url{https://github.com/mraad/}} implements an approximation of DBSCAN algorith with faster but a bit worse results. Another work in~\cite{approx} implements an approximation of DBSCAN algorithm which yield better efficiency in the cost of a bit worse clustering results.  

However, a recent benchmark study~\cite{neukirchen2016survey} shows that MPI based distributed implementations of DBSCAN, e.g., PDSDBSCAN, outperform other Spark implementations~\cite{Irving:15,Litouka}
For MPI based parallel DBSCAN implementations, many existing methods use master-slave model~\cite{Chen:2010,Massimo:2002,xu:2002,brecheisen:2006,fu:2011}. 
In the master-slave mode, the data is partitioned into the slaves, each of which clusters the local data and sends to a master node to merge. The master node sequentially merges all the local clusters to obtain the clustering result. This method has a high communication overhead which makes it inefficient in the merging stage. PDSDBSCAN proposed by Patwary et al.~\cite{Patwary:2012} uses a fully distributed parallel algorithm that employs the disjoint-set structure to speed up communication process. Since data points within the same cluster may be distributed over different workers which result in several disjoint-sets, merging them incurs significant communication costs. 

Another work~\cite{hpdbscan:15} proposes to use a more scalable approach based on a grid-based data index pre-processing, in which data index are resorted and neighbor data points are assigned to the same processor to reduce communication cost. 
Different from that work, in our proposed algorithm, we employ a fast global union approach based on parameter server framework to union the disjoint-sets to alleviate the communication burden. Our method does not require specific data pre-processing and is communication efficient compared to the competing MPI based DBSCAN methods.

\section{Conclusions}
We presented a communication efficient parallel DBSCAN based on Parameter Server, named PS-DBSCAN. This algorithm uses a disjoint-set data structure from~\cite{Patwary:2012} and employed a fast global union approach to union the disjoint-sets to alleviate the communication burden. We compared the performance of PS-DBSCAN with the MPI implementation PDSDBSCAN-D on Real-world datasets and synthetic datasets with different scales. Experiments show that PS-DBSCAN outperforms the MPI-based PDSDBSCAN-D with 2-10 times speedup on communication efficiency in both real-world and synthetic datasets, and the speedup increases with the number of processor cores and the dataset scale. It is shown that combining multithreading into distributed memory system can bring more speedup, we plan to employ multithreading in PS-DBSCAN to further boost the overall efficiency in future.

We have released our PS-DBSCAN in an algorithm platform called Platform of AI (PAI) in Alibaba Cloud and also demonstrated how to use it in PAI.

\begin{acks}
The authors would like to thank Jun Zhou, Xu Chen and Ang Wang for providing valuable comments and helpful suggestions. The authors would also like to thank anonymous reviews for their valuable comments. 
	
	
\end{acks}

\bibliographystyle{ACM-Reference-Format}
\bibliography{sigproc} 

\end{document}